\documentclass[amsmath,nofootinbib,notitlepage,prd,superscriptaddress,twocolumn]{revtex4-2}
\usepackage[T1]{fontenc}
\usepackage[svgnames]{xcolor}
\usepackage{aas_macros,graphicx,hyperref,orcidlink,booktabs,mathrsfs,amsmath,amssymb,physics, bm,placeins}

\hypersetup{colorlinks=true,citecolor=RoyalBlue,linkcolor=RoyalBlue,urlcolor=RoyalBlue}

\begin{document}

\title{Parameter-estimation bias induced by transient orbital resonances in extreme-mass-ratio inspirals}

\author{Edoardo Levati\,\orcidlink{0009-0002-5316-3762}} 
\email{levae25@wfu.edu}
\affiliation{Department of Physics, Wake Forest University, Winston-Salem, North Carolina 27109, USA}
\author{Alejandro C\'ardenas-Avenda\~no\,\orcidlink{0000-0001-9528-1826}} 
\affiliation{Department of Physics, Wake Forest University, Winston-Salem, North Carolina 27109, USA}

\begin{abstract}
Given the multifrequency nature of relativistic orbits, transient orbital resonances are expected to be ubiquitous during an extreme-mass-ratio inspiral (EMRI). At a resonance, the orbital dynamics is modified in a nontrivial way, imprinting an overall dephasing in the emitted gravitational waves and potentially impacting both the detection and parameter estimation of these sources. In this work, using a Fisher-matrix approach, we investigate the bias induced by transient orbital resonances in EMRI parameter estimation. We focus on the most dynamically significant low-order resonances, $3:2$ and $2:1$, as well as on the high-order, subdominant resonances $3:1$ and $4:3$. We find that, for most of the orbits considered, neglecting the effect of a resonance crossing leads to significant losses in signal-to-noise ratio and induces bias in parameter recovery. Furthermore, both the sign and the amplitude of the resonance-induced modifications to the integrals of motion play a crucial role and must be modeled accurately. Our results provide further evidence that failing to model transient orbital resonances accurately can hinder EMRI detection and parameter estimation, thereby limiting their scientific potential.
\end{abstract}

\maketitle

\section{Introduction}

Space-based gravitational wave (GW) detectors, such as the Laser Interferometer Space Antenna (LISA)~\cite{LISA:2024hlh}, will open a new frequency window centered in the mHz band that is beyond the reach of ground-based interferometers. These detectors will be sensitive to new astrophysical sources and augment the information we have so far from ground-based observations. One of the most promising targets is the class represented by~\emph{extreme-mass-ratio inspirals} (EMRIs), in which a stellar-mass compact object of mass $\mu$ spirals into a supermassive black hole (BH) of mass $M$, with mass-ratio $\eta\equiv\mu/M \lesssim 10^{-4}$~\cite{LISA:2022yao}.

Because of their slow, adiabatic evolution, EMRIs are expected to complete around $10^4$–$10^5$ orbital cycles prior to merger~\cite{Hughes,Amaro-Seoane:2007osp,Barack:2003fp}, allowing them to potentially stay in LISA's sensitivity band for its entire mission duration. Estimates indicate that LISA could detect up to $10^3$ EMRI events over its nominal $4.5$-year mission lifetime~\cite{Kludge,Babak:2017tow,LISA:2024hlh}. The GW signal produced by the slow inspiral encodes detailed information about the spacetime geometry of the central object, enabling extremely precise measurements of BH masses and spins~\cite{Ryan:1995wh,Barack:2003fp,Speri:2026ade}, as well as tests of consistency with the predictions of general relativity~\cite{Ryan:1997hg,CardenasSopuertaReview,Speri:2024qak}. Moreover, EMRIs can act as high-precision standard sirens, providing independent constraints on cosmological parameters~\cite{MacLeod:2007jd,Babak:2010ej,Laghi:2026ujb}. Observing several EMRI events will also probe the population of massive BHs and stellar-mass compact objects in the galactic nuclei~\cite{Gair:2004iv,Qunbar:2024prl}.

As for other gravitational-wave sources, EMRI signals are expected to be searched for and characterized with matched-filtering techniques based on large waveform-template banks. In practice, however, their long duration and intricate morphology make fully coherent template-based searches and parameter estimation particularly challenging~\cite{Speri:2025ucn}. Recent work has therefore explored several approaches, including simulation-based inference~\cite{Cole:2025sbi}, semicoherent single-harmonic searches~\cite{Speri:2025ucn}, and full Bayesian inference pipelines~\cite{Speri:2024qak}. Regardless of the inference strategy adopted, accurate and computationally efficient waveform models that capture the relevant EMRI phenomenology across the full parameter space remain essential to fully exploit the scientific potential of space-based GW observations~\cite{Vallisneri:2013rc,Gair:2008ec,Chua:2020stf,Katz:2021yft,LISAConsortiumWaveformWorkingGroup:2023arg}.

EMRI signals can encode a variety of subtle effects, including the spin of the secondary, glitches, tidal resonances induced by nearby perturbers, and signatures of new fundamental fields, which further illustrates the richness of the problem and the need for accurate source modeling~\cite{Piovano2021,boumerdassi2026,BongaYangHughes2019,Maselli_2022}. In this work, we focus on transient orbital resonances as one particularly important and generic effect, and study how neglecting them biases EMRI parameter estimation.

Transient orbital resonances are not an exotic add-on to EMRI dynamics; rather, they arise naturally because, as an inspiral slowly sweeps through Kerr geodesics, the radial and polar frequencies generically pass through low-order rational ratios. These resonances are expected to be ubiquitous during the observationally relevant stage of an EMRI, and can persist for several orbital cycles, making them relevant for waveform modeling and parameter inference schemes~\cite{FlanaganHinderer2,Ruangsri:2013hra,vandeMeent:2013sza,Berry}. 

Following the identification of transient resonances in the seminal work by Flanagan and Hinderer~\cite{FlanaganHinderer2}, subsequent work has developed several complementary strategies to model them. In Ref.~\cite{FlanaganHughes}, using Teukolsky-equation-based calculations, it was shown that on-resonance fluxes depend on the relative phase between the radial and polar motions, providing a strong-field perturbative picture for why resonances can either enhance or suppress the fluxes. Then in Ref.~\cite{Ruangsri:2013hra}, exact Kerr geodesics together with an approximate inspiral model were used to survey the parameter space, showing that low-order resonances are generically encountered and can persist for hundreds of orbital cycles in representative EMRIs. An evolution scheme with discrete resonance-induced modifications to the integrals of motion was introduced in Ref.~\cite{Berry} and used in population studies, finding that many systems cross low-order resonances, though the impact on detectability is mitigated for typical low-eccentricity inspirals. More recently, Ref.~\cite{Lynch:2024ohd} treated resonances within fast self-forced inspirals by switching from a fully averaged evolution to a partially averaged one near resonance, where the slow resonant phase is evolved explicitly. In this work, we follow the approach proposed in Ref.~\cite{SperiGair}, where an effective resonance model is used as a phenomenological surrogate for the same underlying physics: it augments an adiabatic inspiral with a localized, smooth, phase-coherent modification of the fluxes, while remaining informed by the strong-field structure seen in Teukolsky-based calculations.

Building on Ref.~\cite{SperiGair}, which focused on the $3:2$ resonance with fiducial coefficient choices, and Ref.~\cite{Levati}, which used orbit-dependent modifications to the integrals of motion, we use a Fisher-matrix approach to assess the parameter bias induced by transient orbital resonances with orbit-dependent coefficients and representative sign assignments. In particular, we focus on the most dynamically significant \emph{low-order} resonances, the $3:2$ and the $2:1$, as well as on the \emph{high-order}, subdominant resonances, the $3:1$ and the $4:3$. We find that, for most of the orbits we study, spanning low to moderate eccentricities and inclinations, neglecting the effect of a resonance crossing induces a significant bias in parameter estimation.

The rest of this paper is organized as follows. In Sec.~\ref{sec::ERM}, we review the phenomenological model used to investigate the impact
of transient orbital resonances in EMRIs, together with the tools implemented for data analysis purposes. In Sec.~\ref{sec::FM}, we briefly summarize the Fisher matrix formalism and describe how we validate the Fisher analysis. In Sec.~\ref{sec::Results}, we compute the parameter bias induced by different resonances for specific orbits, spanning low to moderate eccentricities and inclinations. In Sec.~\ref{sec::discussion}, we discuss the implications of our findings and outline directions for future work.

\section{Modeling transient orbital resonances in EMRIs}\label{sec::ERM}

EMRI systems experience a slow, adiabatic change in the orbital parameters during their evolution. The fundamental orbital frequencies, radial ($\omega_r$), polar ($\omega_\theta$), and azimuthal ($\omega_\phi$), vary over time, leading the system to cross resonances~\cite{FlanaganHinderer1,Schmidt:2002qk,Fujita:2009us,Ruangsri:2013hra}. A transient orbital resonance occurs when the polar and radial frequencies become commensurate~\cite{Contopoulos,FlanaganHinderer2}, i.e., $l^*\omega_{r}-m^*\omega_{\theta} = 0$, with $l^*$, $m^*$ $\in \mathbb{Z}$, and we denote the resonance by $\omega_\theta/\omega_r = l^*:m^*$. When the system enters the resonance, the evolution of the integrals of motion gains a ``kick,'' due to additional contributions to the adiabatic fluxes, and the trajectory deviates from the adiabatic one~\cite{FlanaganHinderer2,FlanaganHughes}. This behavior essentially happens because the terms in the self-force that would ordinarily average away can add coherently over the resonance time, producing a phase-dependent modification in the constants of motion and, consequently, a corresponding dephasing of the waveform. 

To account for resonant orbits, one must either conduct theoretically intensive gravitational self-force calculations, or find a way to enhance any kludge scheme. The latter approach was pursued in Ref.~\cite{SperiGair}, where a phenomenological model, the effective resonance model (ERM), has been introduced to investigate the impact of transient orbital resonances in EMRIs. The ERM evolves an adiabatic inspiral in the Kerr spacetime using the \textit{numerical kludge}~(NK) model~\cite{Kludge} and, near orbital resonances, modifies the fluxes as~\cite{SperiGair}
\begin{equation}
    \label{Eq:FluxMod}
    \frac{dJ_i}{dt}=f_\mathrm{NK}\left[1+\mathcal{C}_i\,w(t)\right],
\end{equation}
i.e., by an activating impulse function $w(t)$ and adding extra terms to the post-Newtonian fluxes $f_\mathrm{NK}$~\cite{Gair}, with $J_i = (\mathcal{E}, \mathcal{L}_z, \mathcal{Q})$. We refer the reader to Ref.~\cite{Levati} for details about the ERM implementation. The resonance's impact on the orbital evolution is parametrized by the \emph{resonance coefficients} $\mathcal{C}_i = (\mathcal{C}_{\mathcal{E}}, \mathcal{C}_{\mathcal{L}_z}, \mathcal{C}_{\mathcal{Q}})$. Given the functional form of the impulse function~\cite{SperiGair}, the transition between the resonance regime and the adiabatic regime is smooth rather than sharp, in accordance with Figure 1 of Ref.~\cite{Ruangsri:2013hra}. 

In this work, we study the orbital configurations reported in Table~\ref{Table_1}, for which the resonance coefficients are computed from Teukolsky-based calculations~\cite{FlanaganHughes}. These coefficients represent the peak-to-trough variation in the fluxes, as the exact magnitude of the fluxes at resonance varies with the relative phase of the radial and angular motions~\cite{FlanaganHinderer2, Berry, FlanaganHughes}.

As we are interested in studying the phenomenology of different types of resonances, instead of modeling the full resonant-phase dependence, we adopt a reduced phenomenological prescription in which we retain only representative relative-sign patterns among the resonant coefficients. The phase in $\mathcal{Q}$ can be offset with respect to those in $\mathcal{E}$ and $\mathcal{L}_z$, as illustrated in Figure 5 of Ref.~\cite{Berry}; therefore, when crossing a resonance, an EMRI is expected to experience a nonvanishing modification in at least one constant of motion~\cite{Berry}. Consequently, we adopt a prescription consistent with Refs.~\cite{Berry,FlanaganHughes} and evolve different configurations where the modifications in the constants of motion are either all in phase, i.e., $\mathrm{sgn}(\mathcal{C}_{\mathcal{E}}$, $\mathcal{C}_{\mathcal{L}_z}$, $\mathcal{C}_{\mathcal{Q}})$ $=$ ($-$, $-$, $-$) / ($+$, $+$, $+$), or when $\mathcal{E}$ and $\mathcal{L}_z$ are in phase while $\mathcal{Q}$ is out of phase, i.e., $\mathrm{sgn}(\mathcal{C}_{\mathcal{E}}$, $\mathcal{C}_{\mathcal{L}_z}$, $\mathcal{C}_{\mathcal{Q}})$ $=$ ($-$, $-$, $+$) / ($+$, $+$, $-$). 

Under this model, the orbital dephasing caused by a resonance crossing scales approximately as~\cite{Levati}
\begin{equation}\label{dephaseScaling}
    \Delta \Phi_i \sim \mathcal{C} \: \Biggl ( \frac{T}{t_{\textnormal{res}}} \Biggr )^2, 
\end{equation}
where $T$ is the observation time, $t_{\textnormal{res}}$ is the resonance duration, and $\mathcal{C}$ is a scalar proportional to the resonance strength that captures the effect of the resonance-induced modifications to the integrals of motion. In Eq.~\ref{dephaseScaling}, the inverse dependence on $t_{\textnormal{res}}$ reflects the fact that resonances occurring farther from the central black hole last longer but take place in a region where the GW fluxes are weaker~\cite{SperiGair,Levati}. The shift in the orbital phases leads to an overall dephasing in the emitted GWs, and can potentially affect the detection and parameter estimation of EMRI systems~\cite{FlanaganHinderer2, SperiGair, Lynch:2024ohd, Levati}. 

\begin{figure}[]
\includegraphics[width=\columnwidth]{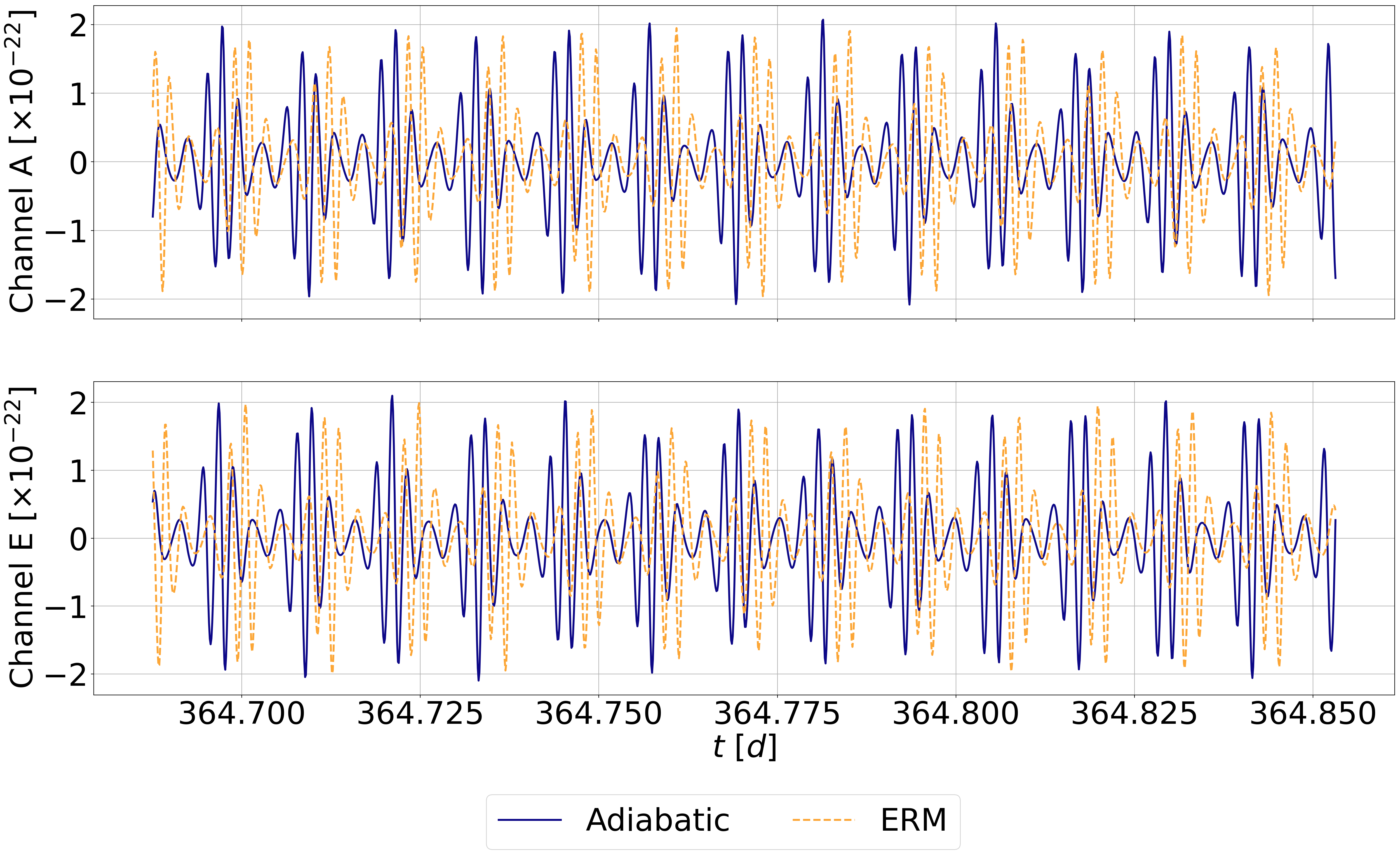}
\caption{The last portion of the waveforms for the EMRI \textit{case (ii)} of Table~\ref{Table_1}, with ($a = 0.9$, $\eta = 10^{-5}$, $M = 10^6 M_{\odot}$, $p/M$ = $8.67$, $e$ = $0.30$, $\iota$ = $1.22$). We use the \texttt{FASTLISARESPONSE}~\cite{Katz:2022yqe} code to project the time-domain signals onto the LISA second-generation time-delay interferometry and construct the $\{A,E\}$ orthogonal channels (top and bottom panels, respectively), making use of the European Space Agency’s simulations of LISA orbits~\cite{lisatools}. The blue solid line corresponds to an adiabatic evolution without accounting for resonance effects. The yellow dashed line corresponds to an evolution where we implement the effective resonance model~\cite{SperiGair}, activating the $3:2$ resonance with the coefficients provided in Ref.~\cite{FlanaganHughes}. We compute the loss in the recovered SNR and the mismatch between the two waveforms, finding $\rho_{\text{eff}}/\rho_{\text{opt}} = 0.24$ and $\mathcal{M} = 7.62 \times 10^{-1}$ at the end of the observation time, $T \approx 1$ year.}
\label{fig1}
\end{figure}

Once the EMRI inspiral has been evolved with the NK or the ERM model, we construct the gravitational waveform by applying the quadrupole formula in the transverse-traceless gauge to the Boyer--Lindquist coordinates of the inspiralling object as a function of the coordinate time $t$. We obtain the two GW polarizations, $h_+(t)$ and $h_{\times}(t)$, and use the \texttt{FASTLISARESPONSE}~\cite{Katz:2022yqe} code to project the time-domain signals onto the LISA second-generation time-delay interferometry (TDI) $\{A,E,T\}$ orthogonal channels, making use of the European Space Agency’s simulations of LISA orbits~\cite{lisatools}. In our analyses, we neglect the $T$ channel because, in the low-frequency regime relevant for EMRIs, it is only weakly sensitive to GW signals~\cite{Prince:2002hp}. Figure~\ref{fig1} shows an example of the waveforms computed for the EMRI \textit{case (ii)} listed in Table~\ref{Table_1}. 

We then apply an overall scaling factor to fix the optimal signal-to-noise ratio (SNR) to $\rho_{\text{opt}} = 20$, usually set as a detection threshold for EMRIs~\cite{Babak:2017tow,Chua:2017ujo}. We define the matched-filter SNR with the noise-weighted inner product~\cite{Cutler:1994ys}; for the LISA TDI $\{A,E\}$ channels, the total SNR is given by $\rho^2 = \rho_A^2 + \rho_E^2$~\cite{Prince:2002hp}. We model the stationary instrumental noise together with the unresolved Galactic-binary background using \texttt{LISATOOLS}~\cite{lisatools} and assume identical, uncorrelated noise power spectral densities for the two channels, i.e., $S_n^A(f) = S_n^E(f)$. 

Lastly, to compare resonant and nonresonant crossing waveforms, we use two standard metrics: (i) the loss in the recovered SNR, quantified by the ratio $\rho_{\text{eff}}/\rho_{\text{opt}}$ following Eqs.~31 and 32 of Ref.~\cite{Burke:2024pfa}, and (ii) the mismatch, defined from the overlap $\mathcal{O}$~\cite{Cutler:1994ys,Owen:1996dk,Moore:2014lga} as $\mathcal{M} = 1 - |\mathcal{O}|$. 
In the limit of small waveform differences, one has $\mathcal{M} \simeq ||\delta h_\perp||^2/(2\rho^2) \le ||\delta h||^2/(2\rho^2)$~\cite{Owen:1996dk,Moore:2014lga}, where $||\delta h||$ is the norm of the waveform difference. A commonly used criterion for waveform indistinguishability is $||\delta h|| < 1$; for our choice $\rho_{\text{opt}} = 20$, this corresponds to $\mathcal{M} \lesssim 10^{-3}$.

\section{Fisher-Matrix Formalism and Validation}
\label{sec::FM}

We will use the Fisher matrix formalism to estimate both the parameter-measurement precision and the systematic error, or bias, induced by waveform mismodeling~\cite{Vallisneri2008, Gair2013, Vallisneri:2013rc}. Although this approach is conceptually straightforward and widely used in the community~\cite{Babak:2017tow, Burke_2020, Piovano2021, SperiGair, Maselli_2022, boumerdassi2026}, obtaining numerically reliable results for EMRIs is notoriously challenging~\cite{Vallisneri2008}. For this reason, we adopt a more conservative strategy and do not rely on a single Fisher estimate. Instead, we identify a region of numerical stability and report the corresponding range of bias values, all consistent with a standard Fisher analysis. In this section, we briefly review the main ingredients of our implementation, the checks we perform to assess the validity of the resulting biases, and the procedure we use to construct these ranges. 

Let $h\left(t, \bm{\lambda}\right)$ be a waveform depending on a set of source parameters $\bm{\lambda}$. In the high-SNR limit, assuming stationary, Gaussian noise, the likelihood is approximated with a multivariate Gaussian distribution, whose curvature around the local maxima defines the Fisher matrix as
\begin{equation}\label{fisherMatrix}
    \Gamma_{ij} = \left( \partial_i h \middle| \partial_j h \right),
\end{equation} 
where $(\cdot | \cdot)$ is the standard noise-weighted inner product as defined in Ref.~\cite{Cutler:1994ys}. We compute the partial numerical derivatives $\partial_i h$ using the five-point stencil formula
\begin{equation}\label{stencil}
\begin{split}
\dv{h}{x} &= \frac{1}{12\epsilon} \bigl[h(x-2\epsilon) - h(x+2\epsilon) + 8h(x + \epsilon) + \\
          &\qquad - 8h(x - \epsilon) \bigr] + \mathcal{O}(\epsilon^4) ,
\end{split}
\end{equation}
where $\epsilon$ is the perturbation step chosen to compute the finite differences. In this work, we choose a relative perturbation, $\epsilon/\lambda^i$, equal for all the parameters.

The inverse of $\Gamma_{ij}$ yields the covariance matrix, whose diagonal elements correspond to the statistical uncertainties of the waveform parameters, $\Delta \lambda^i = \sqrt{(\Gamma^{-1})^{ii}}$, which scale as $1/\rho$ and represent a lower bound on the achievable parameter-measurement precision~\cite{Vallisneri2008}. The off-diagonal elements of the covariance matrix encode the correlations between parameters.

If the GW signal is described by a waveform $h_{\mathrm{t}}$ but it is recovered with an approximate model $h_{\mathrm{m}}$, a bias in the estimate will be induced. Following Ref.~\cite{Cutler2007}, the total parameter shift is
\begin{equation}
    \delta \lambda^i = \delta \lambda^i_{\mathrm{stat}} + \delta \lambda^i_{\mathrm{bias}} ,
\end{equation}
where 
\begin{equation}
    \delta \lambda^i_{\mathrm{stat}} = (\Gamma^{-1})^{ij} \left(\partial_j h \middle| n \right)
\end{equation}
is the statistical error sourced by the noise realization, which scales approximately as $1/\rho$, and
\begin{equation}\label{bias}
     \delta \lambda^i_{\mathrm{bias}} = (\Gamma^{-1})^{ij} \left(\partial_j h \middle| h_{\mathrm{t}} - h_{\mathrm{m}} \right)
\end{equation}
is the systematic error sourced by waveform mismodeling, in our case by neglecting the effect of a resonance crossing, which is independent of the SNR~\cite{Vallisneri:2013rc}. We work in the zero-noise approximation and consider resonance effects to be relevant when the induced bias is larger than the corresponding statistical fluctuation, i.e., 
\begin{equation}\label{ratio}
    \bigg| \frac{\delta \lambda^i_{\mathrm{bias}}}{\Delta \lambda^i} \bigg| > 1.
\end{equation}

Having revisited the Fisher approach, we now turn to the bias induced in the recovered parameters when an ERM waveform is injected as the signal, and a NK waveform is used as the model. Let us now introduce the EMRI parameter space and specify which parameters we focus on in the analysis. 

The EMRI waveform parameter space is commonly divided into intrinsic and extrinsic parameters~\cite{Barack:2003fp,Kludge}. The intrinsic parameters are $\left\{ M,\, \mu,\, e,\, \iota,\, a,\, p/M,\, \psi_0,\, \chi_0 \right\}$, where $\left(M,\, \mu \right)$ are the detector-frame masses of the central supermassive black hole and the inspiraling stellar-mass compact object, $e$ is the eccentricity, $\iota$ is the orbital inclination, $a$ is the dimensionless spin of the central black hole, $p/M$ is the dimensionless semilatus rectum, and $\left(\psi_0,\, \chi_0 \right)$ are the initial radial and polar orbital phases. The extrinsic parameters are $\left\{ \theta_S,\, \phi_S,\, \theta_K,\, \phi_K,\, \phi_0,\, D_L \right\}$, where $\left(\theta_S,\, \phi_S \right)$ are the sky-location polar and azimuthal angles in ecliptic coordinates, $\left(\theta_K,\, \phi_K \right)$ are the polar and azimuthal angles specifying the initial spin direction of the central black hole in ecliptic coordinates, $\phi_0$ is the initial azimuthal orbital phase, and $D_L$ is the luminosity distance. For all the waveforms computed in this work, we choose the initial orbital phases to be $\left( \psi_0,\, \chi_0, \, \phi_0 \right) = \left( 0,\, \pi/2, \, 0 \right)$, and the other extrinsic parameters to be $\left( \theta_S,\, \phi_S,\, \theta_K,\, \phi_K,\, D_L \right) = \left( 0,\, \pi/2,\, 0,\, 0,\, 200 \, \textrm{Mpc}\right)$.

In principle, a full Fisher analysis should account for the complete parameter space. However, this is numerically challenging for EMRIs because the resulting Fisher matrices are typically ill conditioned~\cite{Vallisneri2008}, with a large condition number $k$, defined as the ratio of the largest to the smallest eigenvalue, so that even small inaccuracies in the numerical derivatives can be greatly amplified when the matrix is inverted. 

Following the approach described in Ref.~\cite{SperiGair}, under the assumption that the secular postresonance phase growth is only weakly degenerate with simple phase offsets, we perform a \emph{conditional Fisher analysis} in which only a restricted six-dimensional intrinsic parameter subset is varied, while all remaining parameters, including the extrinsic parameters and the initial phases, are held fixed. Specifically, we consider
\begin{equation*}
    \bm{\lambda} = \left( \log M,\, \log \eta,\, e,\, \iota,\, a,\, p/M \right) ,
\end{equation*}
where the reparametrization $\left(M, \mu \right) \rightarrow \left(\log M, \log \eta \right)$ significantly reduces $k$, improving the stability of the Fisher matrix. The implications of restricting the analysis to this conditional parameter subset, and its relation to a fully marginalized Fisher treatment, are discussed in details in Appendix~\ref{sec::conditional-fisher-bias}.

We validate the Fisher analysis following the consistency checks described in Refs.~\cite{Vallisneri2008,SperiGair,Piovano2021}:
\begin{itemize}
    \item We perturb each element of the Fisher matrix with a deviation matrix $F_{ij}$, whose elements are drawn from a uniform distribution $U[-10^{-3},\,10^{-3}]$. As a stability check, we compute
    \begin{equation}
    \max_{ij}\left[\frac{\left((\Gamma+F)^{-1} - \Gamma^{-1}\right)^{ij}}{(\Gamma^{-1})^{ij}}\right].
    \end{equation}
    For all the cases studied, we find $\max \left\{\textrm{all configurations}\right\} \sim \mathcal{O}(10^{-3})$.
    \item We compare the overlap predicted by the quadratic Fisher expansion with that obtained from small perturbations of the waveform parameters. Starting from
    \begin{equation}
    \begin{aligned}
    \mathcal{O}\!\left(h(\bm{\lambda}+\bm{\gamma}),\,h(\bm{\lambda})\right)
    \sim 1 - \frac{1}{2\rho^2}\,\gamma^i \Gamma_{ij} \gamma^j ,
    \end{aligned}
    \end{equation}
    we compute the ratio between the left-hand side and the right-hand side, finding a maximum deviation of $\mathcal{O}(10^{-2})$.
\end{itemize}

This first check provides a \emph{lower} bound on the values of $\epsilon$ for which $\Gamma$ remains numerically stable, whereas the second one provides an \emph{upper} bound for which the linear-signal approximation remains valid. Together, these two checks identify, for each configuration, a \emph{range} of acceptable values of $\epsilon$ to be used in Eq.~\ref{stencil} to compute the Fisher matrix, and therefore a corresponding range of acceptable bias estimates.

\section{Resulting Parameter Bias}
\label{sec::Results}

We now have all the ingredients needed to assess the parameter bias induced by neglecting transient orbital resonances. To do so, we consider the orbits of Table~\ref{Table_1} and compute the bias induced by the $4:3$, $3:2$, $2:1$, and $3:1$ resonances from Eqs.~\ref{bias} and \ref{ratio}, taking the approximate waveform $h_{\mathrm{m}}$ to be the NK waveform, whereas the injected signal is the ERM waveform. 

\begin{table*}[]
\scriptsize
\renewcommand\arraystretch{1.0}
\setlength{\tabcolsep}{8pt}
\begin{tabular}{l|c|c|c|c|c|c|c|c}
\toprule
Resonance & $e$ & $\iota$ & $p / \textnormal{M}$ & t$_{\textnormal{res}}$ [d] & T$_{\textnormal{max}}$ [d] & $\mathcal{C}_{\mathcal{E}}[\times 10^{-2}]$ & $\mathcal{C}_{\mathcal{L}_z} [\times 10^{-2}]$ & $\mathcal{C}_{\mathcal{Q}}[\times 10^{-2}]$ \\ 
\hline
\hline
\textit{Case (i)} \\
\hline
4:3 & 0.30 & 0.35 & 7.46 & 5.7 & 361 & 0.001/0.00000005 & 0.001/0.000000005  & 0.006/0.000008  \\
3:2 & 0.30 & 0.35 & 5.36 & 2.3 & 83 & 0.102/0.332  & 0.067/0.116  & 0.208/0.687  \\
2:1 & 0.30 & 0.35 & 3.59 & 0.7 & 10 & 0.131/0.084  & 0.179/0.063  & 0.046/0.203  \\
3:1 & 0.30 & 0.35 & 2.92 & 0.2 & 1 & 0.059/0.028  & 0.070/0.035  & 0.310/0.149  \\
\hline
\textit{Case (ii)} \\
\hline
4:3 & 0.30 & 1.22 & 11.36 & 14.6 & 365 &  0.003 & 0.004 & 0.002 \\
3:2 & 0.30 & 1.22 & 8.67 & 6.5 & 365 & 0.303 & 0.123 & 0.123 \\
2:1 & 0.30 & 1.22 & 6.18 & 2.0 & 52 & 0.004 & 0.080 & 0.002 \\
3:1 & 0.30 & 1.22 & 5.07 & 0.5 & 5 & 0.008 & 0.024 & 0.033 \\
\hline
\textit{Case (iii)} \\
\hline
4:3 & 0.70 & 0.35 & 7.58 & 11.4 & 365 & 0.001 & 0.002 & 0.023 \\
3:2 & 0.70 & 0.35 & 5.50 & 4.7 & 143 & 0.127 & 0.078 & 0.210  \\
2:1 & 0.70 & 0.35 & 3.82 & 1.5 & 29 & 0.167 & 0.067 & 0.357 \\
3:1 & 0.70 & 0.35 & 3.28 &  &  & 0.026 & 0.009 & 0.035 \\
\hline
\textit{Case (iv)} \\
\hline
4:3 & 0.70 & 1.22 & 11.47 & 28.2 & 365 & 0.047 & 0.060 & 0.002 \\
3:2 & 0.70 & 1.22 & 8.80 & 12.4 & 365 & 1.030 & 0.489 & 0.261 \\
2:1 & 0.70 & 1.22 & 6.36 & 3.7 & 84 & 0.662 & 0.270 & 0.494 \\
3:1 & 0.70 & 1.22 & 5.41 & 0.8 & 4 & 0.125 & 0.027 & 0.126 \\
\bottomrule
\end{tabular}
\caption{Parameter-space location and resonance coefficients for the investigated orbital resonances, with $a = 0.9$, $\eta = 10^{-5}$ and $M = 10^6 M_{\odot}$. We consider specific orbits passing through different resonances for which the resonance coefficients were computed from Teukolsky-based calculations~\cite{FlanaganHughes}. For the EMRI \textit{case (i)}, we also report the coefficients computed with the surrogate self-force model detailed in Ref.~\cite{Lynch:2024ohd} (i.e., slash-separated entries are listed as Teukolsky-based and self-force-based), whose approximations tend to significantly underestimate the effects of high-order resonances. The EMRI evolution begins close to the resonance, by setting the initial value of $p/M$ such that the threshold function $\xi = (\omega_{\theta}/\omega_r) - \mathscr{R}$~\cite{SperiGair}, where $\mathscr{R} = \frac{l^*}{m^*}$ denotes the resonance value, is $\xi_0 = - (0.002, 0.002, 0.02, 0.05)$ for the $4:3$, $3:2$, $2:1$ and $3:1$ resonance, respectively. We fix the observation time to $T = 1$ year and terminate the evolution earlier if the secondary reaches the separatrix. The resonance duration, $t_{\textnormal{res}}$, is defined as in Ref.~\cite{SperiGair}. For the $3:1$ resonance, we find that the kludge fluxes, particularly those associated with the Carter constant, break down for the high-eccentricity, low-inclination orbit, which starts very close to the separatrix, yielding a trajectory that is not consistent with an adiabatic evolution. Therefore, in our analyses, we do not consider this specific configuration.}
\label{Table_1}
\end{table*}

As discussed in Sec.~\ref{sec::ERM}, the relative phase of the radial and polar motions determines the exact value of the resonance coefficients, including their sign. Since our goal is to assess whether the induced bias can be significant, rather than to track the full phase dependence of the resonance coefficients, we do not model that dependence explicitly. Instead, following Refs.~\cite{Berry,FlanaganHughes}, we restrict the analysis to the sign of the coefficients and investigate four representative sign configurations: cases in which the modifications to the fluxes are all in phase and cases in which $\mathcal{E}$ and $\mathcal{L}_z$ are in phase while that in $\mathcal{Q}$ is out of phase.

\begin{figure*}[]
\includegraphics[scale=0.103]{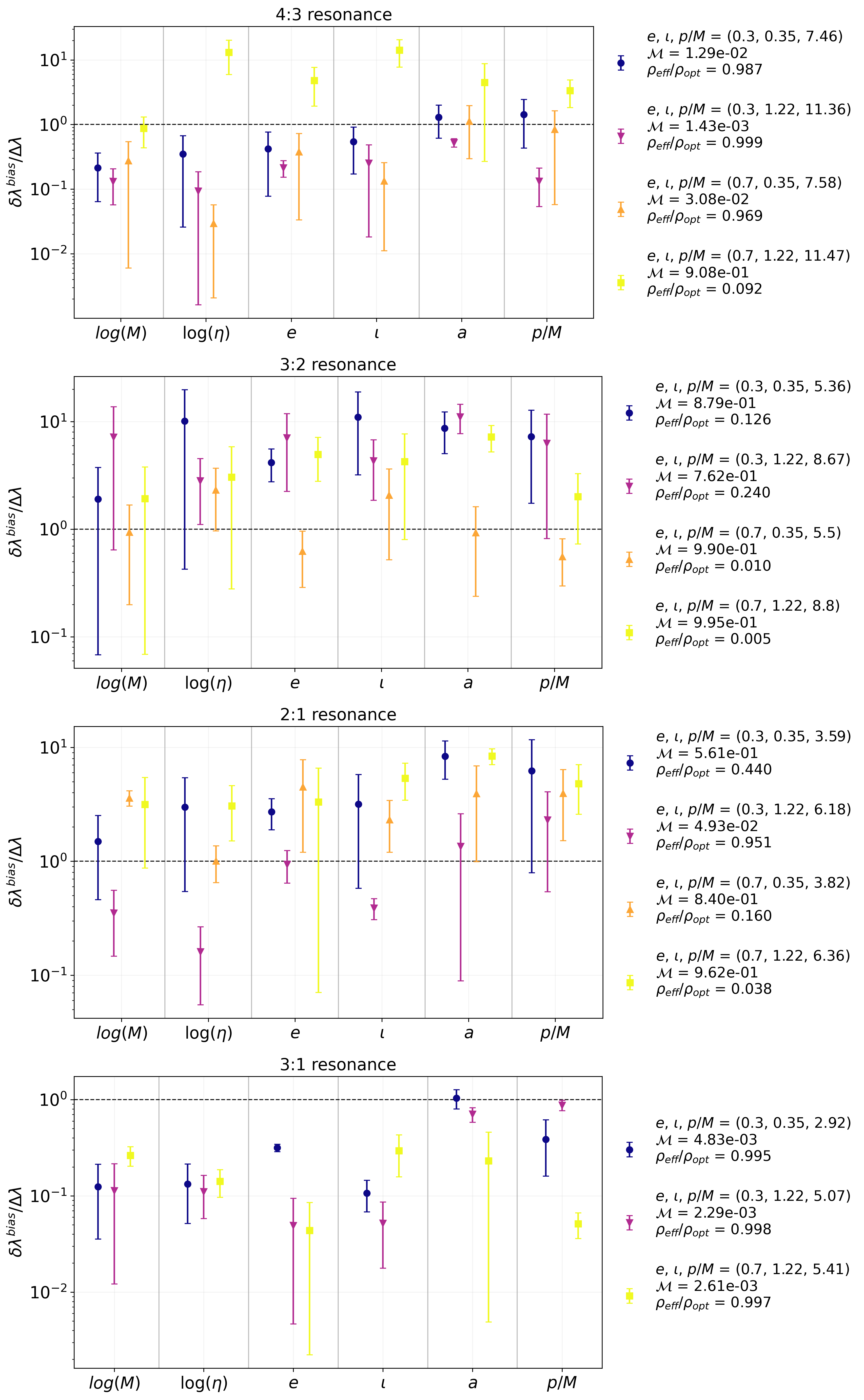} 
\caption{Parameter bias induced by the $4:3$, $3:2$, $2:1$ and $3:1$ resonances, from top to bottom, respectively, for the EMRIs of Table~\ref{Table_1} with $a = 0.9$, $\eta = 10^{-5}$ and $M = 10^6 M_{\odot}$. We use the resonance coefficients provided in Ref.~\cite{FlanaganHughes}, which are computed from Teukolsky-based calculations. We report the results for a specific choice of the signs of the coefficients, i.e., $\mathrm{sgn}(\mathcal{C}_{\mathcal{E}}$, $\mathcal{C}_{\mathcal{L}_z}$, $\mathcal{C}_{\mathcal{Q}})$ $=$ ($-$, $-$, $-$). For each orbit, we compute the loss in the recovered SNR, the mismatch at the end of the observation time between resonant and nonresonant crossing waveforms, and the ratio between the parameter bias and the corresponding statistical uncertainty, from Eq.~\ref{ratio}. These ranges are obtained from the Fisher-validation procedure described in Sec.~\ref{sec::FM}, reflecting the allowed variation in the finite-difference step $\epsilon$. In the case of the $3:1$ resonance, we do not show the results for the high-eccentricity, low-inclination orbit since we find that the kludge fluxes break down, and yield a trajectory that is not consistent with an adiabatic evolution. The horizontal dashed black line marks the threshold  $\left|\delta\lambda_{\textrm{bias}}^{i}\right|/\Delta\lambda^{i}=1$.}
\label{fig2}
\end{figure*}

\begin{figure}[]
\includegraphics[width=\columnwidth]{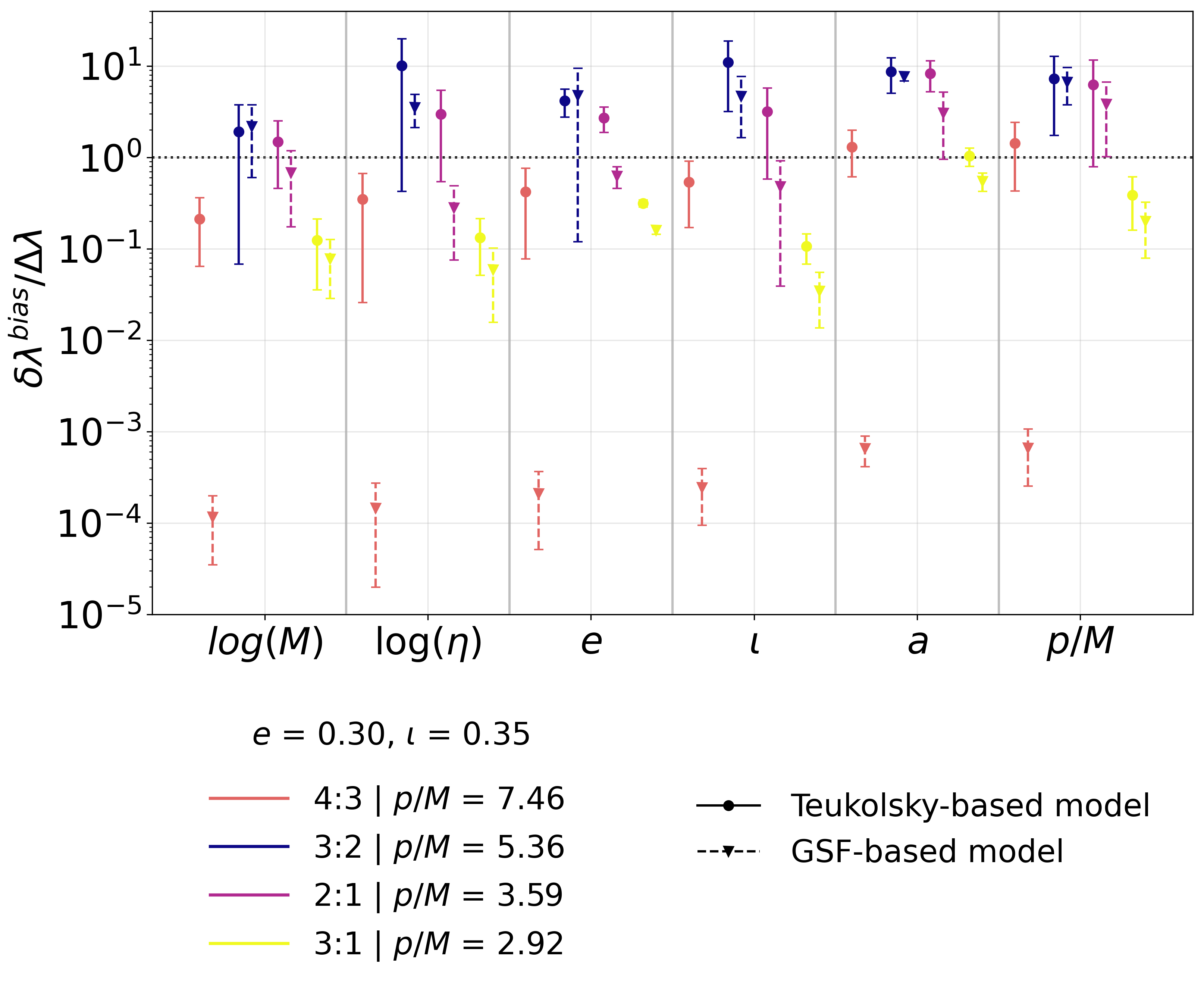}
\caption{Parameter bias for the EMRI \textit{case (i)} of Table~\ref{Table_1}, for which two independent sets of resonance coefficients are available. We compare the bias induced by the $4:3$, $3:2$, $2:1$, and $3:1$ resonances when using the coefficients reported in Ref.~\cite{FlanaganHughes}, obtained from Teukolsky-based calculations, and those computed using a self-force-based model, as provided in Ref.~\cite{Lynch:2024ohd}. The choice of the signs of the coefficients is the same as in Figure~\ref{fig2}. While, for most resonances, the two prescriptions are in overall agreement, they differ significantly in one case, namely the $4:3$ resonance, where the induced bias differs by several orders of magnitude. This behavior is consistent with the findings reported in Appendix A of Ref.~\cite{Lynch:2024ohd}, where the $4:3$ case is identified as a limitation of its self-force-based model, whose approximations tend to significantly underestimate the effects of high-order resonances. Therefore, the figure should be interpreted primarily as a sensitivity test to a known limitation of a surrogate model, rather than as a comparison between Teukolsky-based and full self-force calculations.}
\label{fig3}
\end{figure}

The results shown in Figures~\ref{fig2} and \ref{fig3} refer to the case $\mathrm{sgn}(\mathcal{C}_{\mathcal{E}}$, $\mathcal{C}_{\mathcal{L}_z}$, $\mathcal{C}_{\mathcal{Q}})$ $=$ ($-$, $-$, $-$). Choosing instead $\mathrm{sgn}(\mathcal{C}_{\mathcal{E}}$, $\mathcal{C}_{\mathcal{L}_z}$, $\mathcal{C}_{\mathcal{Q}})$ $=$ ($+$, $+$, $+$) yields the same losses in SNR and mismatches and does not substantially change the overall conclusions of the Fisher analysis. This is because the resonant-crossing waveform is dephased relative to the adiabatic one by the same amount, but in the opposite direction. The horizontal dashed black line in these figures marks the threshold $\left|\delta\lambda_{\textrm{bias}}^{i}\right|/\Delta\lambda^{i}=1$; values above this line indicate that the systematic bias exceeds the corresponding 1$\sigma$ statistical uncertainty.

\begin{figure*}[t]
\includegraphics[width=\textwidth]{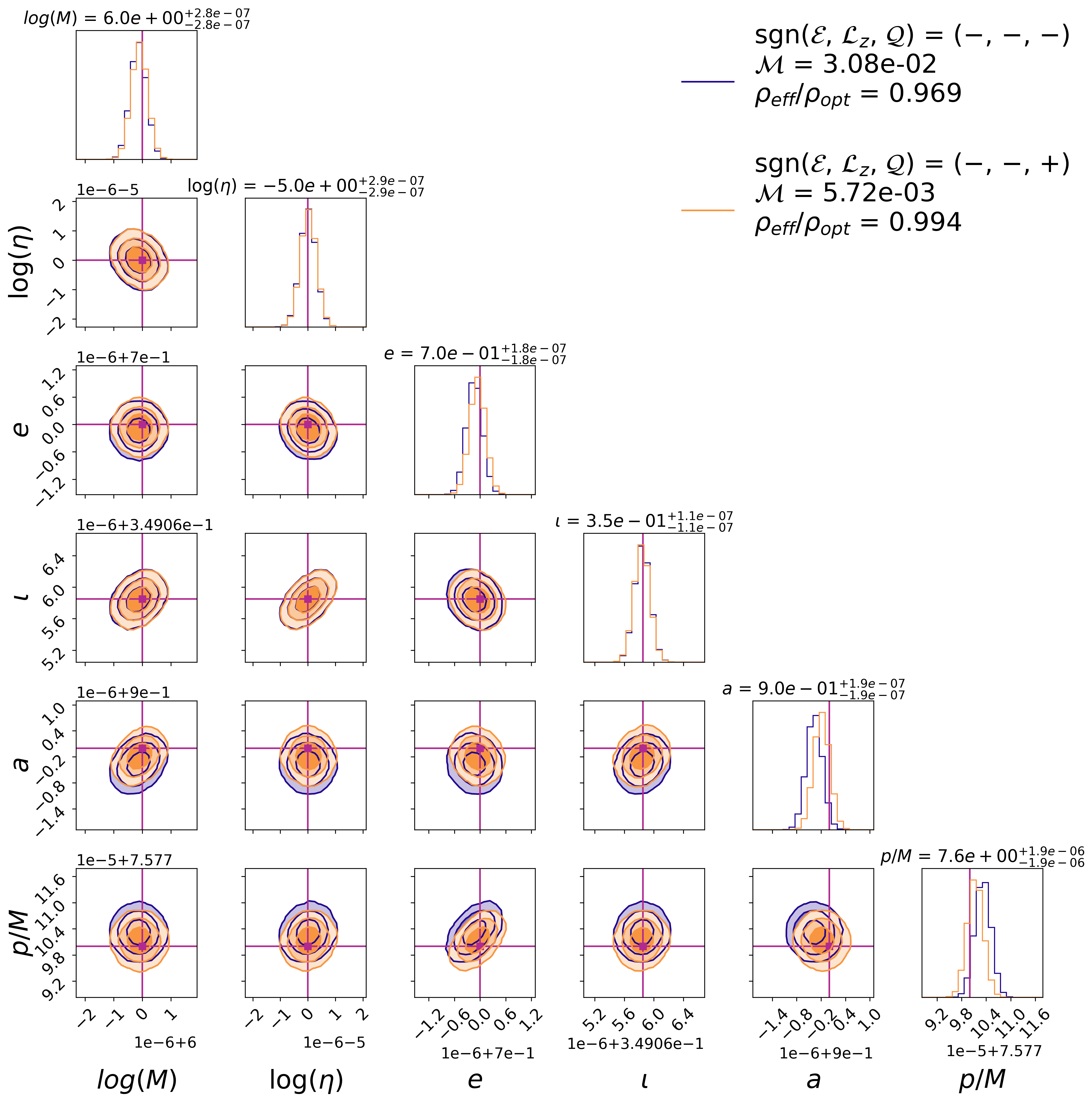} 
\caption{Gaussian posterior approximation from the Fisher matrix for the EMRI \textit{case (iii)} of Table~\ref{Table_1}, where we activate the $4:3$ resonance. The purple line represents the injected parameter values. We compare two different choices for the signs of the resonance coefficients. The blue contours illustrate the induced parameter bias when the modifications to the constants of motion are all in phase, i.e., $\mathrm{sgn}(\mathcal{C}_{\mathcal{E}}$, $\mathcal{C}_{\mathcal{L}_z}$, $\mathcal{C}_{\mathcal{Q}})$ $=$ ($-$, $-$, $-$). The orange contours refer to the case where the modifications in $\mathcal{E}$ and $\mathcal{L}_z$ are in phase, while that in $\mathcal{Q}$ is out of phase, i.e., $\mathrm{sgn}(\mathcal{C}_{\mathcal{E}}$, $\mathcal{C}_{\mathcal{L}_z}$, $\mathcal{C}_{\mathcal{Q}})$ $=$ ($-$, $-$, $+$). For this second choice, the bias induced in $a$ and $p/M$ is lower and falls within the $1\sigma$ threshold.}
\label{fig4}
\end{figure*}

The most relevant sign choice arises when the modification to the Carter-constant flux has the opposite sign, i.e., $\mathrm{sgn}(\mathcal{C}_{\mathcal{E}}$, $\mathcal{C}_{\mathcal{L}_z}$, $\mathcal{C}_{\mathcal{Q}})$ $=$ ($-$, $-$, $+$) or ($+$, $+$, $-$). For the strongest resonances, $3:2$ and $2:1$, changing the sign of $\mathcal{C}_{\mathcal{Q}}$ leads to slightly different numerical results but does not alter the overall conclusions, since in these cases the induced bias is already significantly above $1\sigma$. By contrast, for the weaker resonances, $4:3$ and $3:1$, this sign choice can be decisive, as it can shift the bias from above to below the $1\sigma$ threshold, as illustrated in Figure~\ref{fig4}.  

Overall, we find that for most of the orbits considered neglecting the effect of a resonance crossing leads to significant losses in SNR and induces a systematic error in parameter estimation that exceeds the corresponding statistical fluctuation; i.e., it is larger than $1\sigma$. The $3:2$ and $2:1$ resonances produce the largest biases, although the $4:3$ resonance also exhibits at least one orbit for which resonance effects are significant.  

These results can be understood using the empirical scaling of Eq.~\ref{dephaseScaling}. Some systems plunge within the one-year observation time and accumulate less dephasing after the resonance crossing, leading to a smaller bias. On the other hand, the systematic error is smaller when the resonance is encountered farther away from the central black hole. When those effects are comparable, the bias is primarily driven by the resonance strength, parametrized by the resonance coefficients. 

For the $4:3$ resonance, the orbit with the larger bias is also the one with the largest resonance coefficients, while in the case of the $2:1$ resonance, the orbit with the smaller bias is the one characterized by the smallest values of $\mathcal{C}$. For the $3:1$ resonance, we find that the kludge fluxes, particularly those associated with the Carter constant, break down for the high-eccentricity, low-inclination orbit, which starts very close to the separatrix, yielding a trajectory that is not consistent with an adiabatic evolution. Therefore, we do not report this specific configuration. 

Just by considering the physical parameters entering Eq.~\ref{dephaseScaling}, we cannot explain why, in the case of the $3:2$ resonance, one orbit exhibits lower biases, though reporting a loss in the SNR and a mismatch comparable, if not higher, with respect to the other configurations. Such behavior may indicate either the presence of higher-order contributions beyond the linear-signal approximation~\cite{Vallisneri:2013rc} or a different projection of the resonance-induced dephasing onto the correlated directions of the Fisher matrix for that specific orbit.

The results presented so far are based on the resonant coefficients reported in Ref.~\cite{FlanaganHughes} from Teukolsky-based calculations. As shown for all the orbits studied, the specific values for these coefficients, including their sign, are clearly important in determining the size of the induced bias. We compare against an alternative implementation in which resonance-induced modifications are obtained from a self-force-based model~\cite{Lynch:2024ohd}. Owing to the specific outer-product ansatz used to construct the mixed radial–polar modes, this surrogate model tends to significantly underestimate the effects of high-order resonances. Figure~\ref{fig3} compares the orbits for which both coefficient prescriptions are available. While, for most resonances, the inferred bias is qualitatively similar, it differs significantly for the $4:3$ resonance, in agreement with the limitations of the model discussed in Ref.~\cite{Lynch:2024ohd}. We stress that this figure should be interpreted primarily as a sensitivity test to a known limitation of a surrogate model, rather than as a comparison between Teukolsky-based and full self-force calculations.

Different choices for the signs of the coefficients can affect the Fisher-analysis results, especially for the weaker resonances. In Figure~\ref{fig4}, we compare two configurations, one where the modifications to the constants of motion are all in phase and one where those in $\mathcal{E}$ and $\mathcal{L}_z$ are in phase while that in $\mathcal{Q}$ is out of phase. For this second choice, the bias induced in some parameters, namely $a$ and $p/M$, is lower and falls within the $1\sigma$ threshold. 

\section{Conclusions}\label{sec::discussion}

In this work, we used a Fisher-matrix analysis to study the parameter bias induced by transient orbital resonances in EMRIs. We revisited the $3:2$ resonance, previously analyzed in Ref.~\cite{SperiGair}, and extended the analysis to the $4:3$, $2:1$, and $3:1$ resonances within a common effective-resonance framework, evaluating the corresponding parameter biases for the orbital configurations listed in Table~\ref{Table_1}.

The effective resonance model adopted in this work~\cite{SperiGair} is informed by Teukolsky-based calculations of the resonance-induced modifications to the fluxes~\cite{FlanaganHughes}. In particular, as in Ref.~\cite{Levati}, we used orbit-dependent coefficients and assigned separate values to the modifications associated with the fluxes of $\mathcal{E}$, $\mathcal{L}_z$, and $\mathcal{Q}$. Within this setup, we found that, for most of the orbits considered, neglecting resonance effects produces losses in the recovered SNR, as well as parameter biases above the corresponding statistical uncertainty. For the $3:2$ resonance, the inferred bias is smaller than that reported in Ref.~\cite{SperiGair}, but remains above the $1\sigma$ threshold. 

Since Fisher analyses for EMRIs are numerically sensitive, we did not report a single representative result for each configuration. Instead, for each case, we identified a region in which the Fisher matrix is numerically stable and reported the corresponding range of bias values. This is the procedure used throughout the paper and it sets the interpretation of all quoted Fisher results.

Our results show how both the sign and the magnitude of the resonance coefficients affect the inferred bias. Different sign assignments have the largest impact for the weaker resonances, for which the inferred bias can move across the $1\sigma$ threshold. Different coefficient prescriptions also lead to different bias estimates. Figure~\ref{fig3} compares the Teukolsky-based coefficients used in the main analysis with those reported in Ref.~\cite{Lynch:2024ohd}, which were constructed using the surrogate self-force model described therein. These two prescriptions are broadly consistent for most of the resonances considered, but they yield significantly different bias estimates for the $4:3$ resonance. Such behavior is in agreement with the findings reported in Appendix A of Ref.~\cite{Lynch:2024ohd}, where the $4:3$ case is identified as a limitation of its surrogate model, whose approximations tend to significantly underestimate the effects of high-order resonances.

According to Eq.~\ref{ratio}, the parameter bias is independent of SNR, while the corresponding statistical uncertainty scales as $1/\rho$. Since we fix the SNR to a commonly adopted EMRI detection threshold, the results reported in Figure~\ref{fig2} should be regarded as lower limits only within the conditional Fisher-matrix framework described in Appendix~\ref{sec::conditional-fisher-bias}. Extending the analysis to the complete parameter space modifies both the statistical uncertainties and the induced biases through parameter correlations; hence, these results should not be interpreted as predictions for a fully marginalized parameter-estimation problem. Furthermore, for large mismatches and significant losses in the recovered SNR, the first-order Cutler--Vallisneri bias formula in Eq.~\ref{bias} can no longer be regarded as an accurate estimate of the systematic shift in the likelihood peak. In this regime, the Fisher-matrix validation probes only the local quadratic structure of the model manifold and, in our case, does not guarantee that the NK-ERM discrepancy remains within the regime of validity of the linearized bias approximation. Consequently, large parameter biases inferred from the Fisher analysis should be interpreted primarily as qualitative indicators.

While the \textit{kludge} waveforms used in this work might not meet the requirements for EMRI searches in LISA data~\cite{FEW}, they do capture the main qualitative effect of a transient resonance, namely, a localized modification of the inspiral that produces a cumulative waveform dephasing and, consequently, a bias in parameter recovery. For this reason, the SNRs, mismatches, and parameter biases reported in this work should be interpreted as a characterization of the impact of transient resonances within this approximate modeling framework; more accurate waveform models may change the quantitative values of the inferred biases, but the qualitative conclusion that neglecting transient resonances can bias EMRI parameter estimation is expected to remain valid. 

The present work provides a Fisher-based case study of selected representative EMRIs within a phenomenological effective-resonance framework and for a specific choice of spin and mass ratio. While the qualitative behavior identified here is expected to persist more broadly, a more complete assessment will require extending the available resonance-coefficient calculations beyond the cases considered in this work. Consequently, there are several natural extensions of this work. One would be to compare these results with sampling-based inference methods, such as MCMC analyses~\cite{boumerdassi2026}, in order to test the Fisher predictions and quantify effects beyond the linear-signal approximation of Eq.~\ref{bias}. Likewise, implementing a resonance treatment in \texttt{FEW}~\cite{FEW} and extending the analysis to inspirals that cross multiple resonances~\cite{Levati} would allow the cumulative impact of successive crossings, together with the correlations they induce in parameter recovery, to be studied within a more accurate waveform framework. A population-level study based on astrophysically motivated source distributions, along the lines of Ref.~\cite{Berry}, will also be needed to determine how representative these case studies are and how often resonance-induced biases are expected to be relevant for LISA observations. We leave these developments to future work.

\begin{acknowledgments}
We thank Kyriakos Destounis, Scott Hughes, Lennox Keeble, Philip Lynch, Pablo Ruales, and Carlos Sopuerta for useful comments and discussions. We also thank the anonymous referee for their comments. Computations were performed using the Wake Forest University (WFU) High Performance Computing Facility, a centrally managed computational resource available to WFU researchers, including faculty, staff, students, and collaborators~\cite{WakeHPC}. 
\end{acknowledgments}

\appendix

\section{Conditional Fisher analysis and systematic bias}
\label{sec::conditional-fisher-bias}

The results presented in this work are derived using a conditional Fisher analysis. In this Appendix, we provide a description of this framework and its relation to a fully marginalized Fisher treatment.

Let us consider Eq.~\ref{bias} and split the complete parameter set as $\lambda^i=(\lambda^a,\xi^A)$, where $\lambda^a$ are the parameters varied in the Fisher analysis, whereas $\xi^A$ are the parameters held fixed at their fiducial values. Let $A_{ab}=\Gamma_{ab}$, $B_{aB}=\Gamma_{aB}$, $C_{AB}=\Gamma_{AB}$, where $\Gamma$ is the Fisher matrix from Eq.~\ref{fisherMatrix}, and define the bias source $\beta_i\equiv(\partial_i h_m\mid h_t-h_m)$.

If the parameters $\xi^A$ are kept fixed at $\xi^A_0$, the Fisher matrix is computed from the conditional likelihood
\begin{equation}
    {\cal L}_{\rm cond}(s\mid\lambda^a)={\cal L}(s\mid\lambda^a,\xi^A_0),
\end{equation}
and it characterizes the local curvature of the likelihood on the submanifold $\xi^A = \xi^A_0$. It is therefore distinct from the Fisher matrix obtained after marginalizing over $\xi^A$, which is given by the Schur complement~\cite{Gallier2019Schur, Datta:2020vcj}
\begin{equation}
    S_{ab} = A_{ab} - B_{aA}\left(C^{-1}\right)^{AB}B_{bB}.
\end{equation}

Within a conditional Fisher analysis, the systematic error in the parameters $\lambda^a$ is
\begin{equation}
\delta\lambda^a_{{\rm bias},\,{\rm cond}} = \left(A^{-1}\right)^{ab}\beta_b,
\label{conditional-bias}
\end{equation}
whereas the bias obtained from a full Fisher analysis after marginalizing over $\xi^A$ is
\begin{equation}
\delta\lambda^a_{{\rm bias},\,{\rm full}} = \left(S^{-1}\right)^{ab}\left[\beta_b-B_{bA}\left(C^{-1}\right)^{AB}\beta_B \right].
\label{full-bias}
\end{equation}

Since $S\leq A$, the conditional analysis underestimates the statistical uncertainties, i.e.,
\begin{equation}
    \Delta \lambda^a_{\rm full} = \sqrt{(S^{-1})^{aa}} \geq \sqrt{(A^{-1})^{aa}} = \Delta \lambda^a_{\rm cond}.
\end{equation}
However, Eq.~\ref{conditional-bias} and Eq.~\ref{full-bias} show that there is no monotonic ordering between the conditional bias and the full bias, because both the Fisher operator and the source term change. Hence, the systematic shift in the parameters $\lambda^a$ can become larger, smaller, or even change sign, depending on how $h_t - h_m$ projects onto the omitted-parameter tangent directions.

We therefore emphasize that the Fisher matrix and bias estimates presented in this work are conditional on the subset of parameters included in the analysis. Such estimates provide a useful diagnostic but should not be interpreted as predictions for a fully marginalized parameter-estimation problem. Since extending the analysis to the complete parameter space modifies both the statistical uncertainties and the systematic shifts through parameter correlations, a conditional bias-to-noise ratio exceeding unity does not necessarily imply that the corresponding parameter would remain biased at the $>1\sigma$ level after marginalizing over all parameters.

\bibliography{resonances_induced_bias}

\end{document}